\documentclass[journal]{IEEEtran}
\usepackage{booktabs} 
\usepackage{ifpdf}
\usepackage{url}

\usepackage{xspace}

\usepackage{balance}
\usepackage{caption}

\usepackage[tight,footnotesize]{subfigure}
\usepackage{multirow}
\usepackage{color}
\usepackage{algorithmicx}
\usepackage{comment}
\usepackage{algorithm}
\usepackage{algpseudocode}

\floatname{algorithm}{Algorithm}

\usepackage{makecell} 

\usepackage{mathtools}

\newcommand{\ie}{\textit{i}.\textit{e}.\xspace}

\begin{document}

\title{CLEAR: \underline{C}hannel \underline{L}earning and \underline{E}nhanced \underline{A}daptive \underline{R}econstruction for Semantic Communication in Complex Time-Varying Environments}
\author{
Hongzhi~Pan,
Shengliang~Wu,
Lingyun~Wang,
Yujun~Zhu,
Weiwei~Jiang,
Xin~He,~\IEEEmembership{Member,~IEEE}
\thanks{This work has been in part supported by Anhui Provincial Higher Education Research Project-Key Project under grant No. 2023AH052300, and the Natural Science Foundation of China under grant No. 62072004. {\it H. Pan and S. Wu are co-primary authors, X. He is the corresponding author.}}
\thanks{H. Pan is with the School of Information and Artificial Intelligence, Anhui Business College, Wuhu, 241002, Anhui, China (email: panhongzhi@abc.edu.cn).}
\thanks{W. Jiang is with the School of Computers, Nanjing University of Information Science and Technology, Nanjing, 210044, Jiangsu, China (e-mail: weiwei.jiang@nuist.edu.cn).}
\thanks{S. Wu, L. Wang, Y. Zhu, and X. He are with the School of Computer and Information, Anhui Normal University, Wuhu, 241002, Anhui, China (e-mail: \{xin.he, zhuyujun, lingyunwang, wslwsl\}@ahnu.edu.cn).}
}

\maketitle
\begin{abstract} 
To address the challenges of robust data transmission over complex time-varying channels, this paper introduces channel learning and enhanced adaptive reconstruction (CLEAR) strategy for semantic communications. CLEAR integrates deep joint source-channel coding (DeepJSCC) with an adaptive diffusion denoising model (ADDM) to form a unique framework. It leverages a trainable encoder-decoder architecture to encode data into complex semantic codes, which are then transmitted and reconstructed while minimizing distortion, ensuring high semantic fidelity. By addressing multipath effects, frequency-selective fading, phase noise, and Doppler shifts, CLEAR achieves high semantic fidelity and reliable transmission across diverse signal-to-noise ratios (SNRs) and channel conditions. Extensive experiments demonstrate that CLEAR achieves a 2.3~dB gain on peak signal-to-noise ratio (PSNR) over the existing state-of-the-art method, DeepJSCC-V. Furthermore, the results verify that CLEAR is robust against varying channel conditions, particularly in scenarios characterized by high Doppler shifts and strong phase noise.
\end{abstract}


\IEEEpeerreviewmaketitle
\hfill\par

\section{Introduction}
The increasing demand for high-quality data transmission in wireless communication systems is challenging due to the dynamic and unreliable nature of channel environments. Traditional communication systems, which prioritize exact data recovery at the bit level, often fail to maintain performance under impairments such as multipath effects, frequency-selective fading, etc. These limitations become particularly evident in scenarios like mobile networks and the Internet of Things (IoT), where channel conditions fluctuate rapidly and unpredictably.

Semantic communication~\cite{Niu2022, Yang2022} is gaining momentum, focusing on conveying the meaning of information rather than ensuring accurate data transmission. This approach is particularly valuable in wireless communication systems, where channel conditions are often dynamic and unreliable. Unlike traditional communication systems, which prioritize bit-level accuracy, semantic communication aims to preserve the intended meaning of transmitted data, even in the presence of channel impairments such as multipath effects, frequency-selective fading, phase noise, and Doppler shifts. These challenges are prevalent in mobile networks, and the Internet of Things, driving the need for adaptive strategies which can maintain communication performance in these environments.
\begin{figure*}[t]
    \centering
    \includegraphics[width=6in]{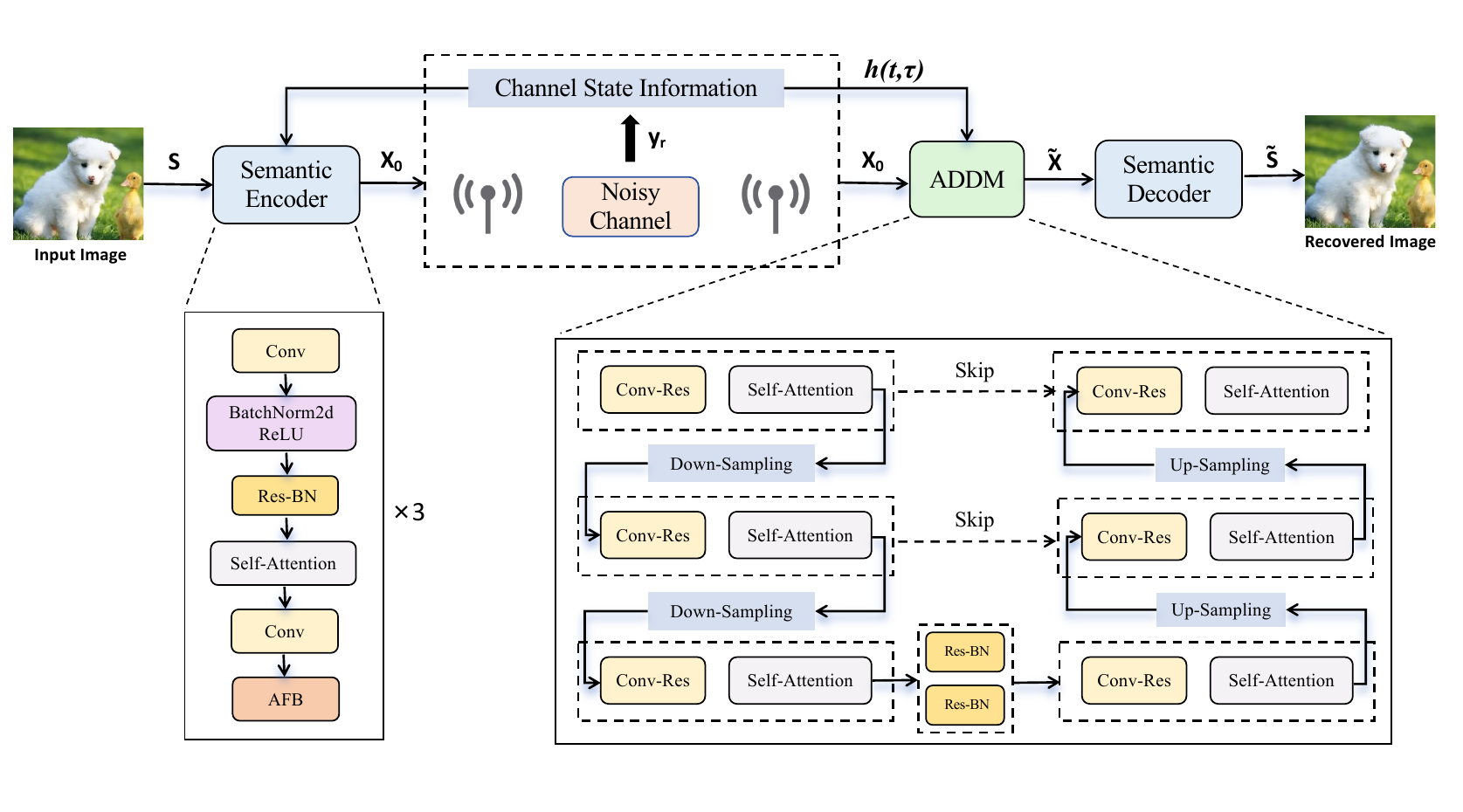}
    \caption{The architecture of CLEAR. The framework integrates DeepJSCC with ADDM, using real-time CSI for dynamic adaption. The encoder extracts semantic features, the channel introduce noise, and the ADDM mitigates distortion to ensure high-quality reconstruction by the decoder.}
    \label{fig:sys_model}
\end{figure*}

In traditional communication systems, source and channel coding is treated as separate processes, based on Shannon's source-channel separation theorem~\cite{Cover2006}. However, this modular approach, which optimizes each coding method independently, does not adequately leverage the interaction between them, leading to suboptimal performance. For example, under low signal-to-noise ratio (SNR) conditions, channel decoders often cannot achieve error-free reconstruction, resulting in data degradation. This issue, known as the {\it cliff effect}~\cite{Gunduz2008cliffeffect}, highlights the limitations of independent source-channel coding.

Therefore, joint source-channel coding (JSCC)~\cite{Zhai2005JSCC, Kostina2017JSCC, Shao2024JSCC} has been proposed, aiming to optimize the coding process at the system level by integrating the design and optimization of source and channel coding. Recent advances in deep learning (DL) have enabled DL models to achieve strong data compression and noise resistance capabilities, implementing end-to-end JSCC processes through a single model. Recent studies have demonstrated that DeepJSCC~\cite{Weng2024DeepJSCC, Lyu2024DeepJSCC}, particularly those based on deep autoencoders, achieve remarkable performance in data compression and transmission reliability across a wide range of wireless communication tasks.

However, several challenges persist in the field of semantic communication using DeepJSCC. One major issue is channel adaptability, i.e., most DeepJSCC models are trained for specific channel conditions, which makes them less effective in highly dynamic environments. Additionally, noise mitigation remains a key challenge, as complex channel impairments, such as Doppler shifts and phase noise, require sophisticated denoising mechanisms to preserve semantic integrity. Lastly, achieving end-to-end optimization for real-time adaptability, while maintaining semantic fidelity across varying SNRs, continues to be an open problem in the field.

To address these challenges, we propose CLEAR, a novel framework for adaptive semantic communication in complex, time-varying channels. Our approach integrates DeepJSCC with an adaptive diffusion denoising model (ADDM) to construct a robust encoder-decoder system. In the CLEAR system, the encoder transforms raw data into meaningful semantic features, which are transmitted over wireless channels. Upon transmission, the features undergo diffusion noise injection followed by sampling-based denoising through ADDM, effectively recovering semantic features with acceptable distortions. Finally, the decoder reconstructs the original information. The primary objective of the process is to minimize semantic loss. A key innovation of our method is the dynamic adjustment of noise parameters based on real-time channel state information (CSI)~\cite{Xie2020CSI}, which effectively mitigates the adverse effects of multipath propagation, frequency-selective fading, phase noise, and Doppler shifts.

The main contributions of this work are summarized as follows. 
\begin{itemize}
    \item {\it Integration of DeepJSCC and ADDM.} By combining joint source-channel coding with advanced denoising capabilities, CLEAR enables efficient encoding and decoding of semantic data. This integration not only eliminates the need for separate encoding steps but also provides the system with end-to-end adaptive capabilities, thereby enhancing data compression efficiency and noise resilience across varying environmental conditions.
    \item {\it Dynamic Adaptation Based on CSI.} Dynamic adaptive CSI enables CLEAR to adjust noise parameters according to varying channel conditions, effectively mitigating the adverse effects of multipath propagation, fading, and noise. This capability allows the system to handle complex channel impairments, such as Doppler shifts and phase noise, ensuring consistent communication quality across diverse environments.
    \item  {\it Extensive validations.} Our experimental results demonstrate CLEAR's superior performance compared to existing approaches, particularly in its ability to preserve semantic fidelity and achieve high-quality data reconstruction, even under challenging conditions. 
\end{itemize}

The rest of the paper is structured as follows. In Section II, related work is presented. Section III presents the system model, detailing the encoder-decoder architecture and the role of CSI in parameter estimation. Section IV provides an in-depth discussion of ADDM, including its diffusion and sampling algorithms, and outlines the training algorithm for the CLEAR system. Section V presents experimental results, demonstrating the effectiveness of our approach in various communication scenarios. Finally, Section VI concludes the paper and suggests potential directions for future research.

\section{Related Work}
In traditional methods, images are typically compressed using JPEG/JPEG2000~\cite{Marcellin2000JPEG}, followed by channel coding techniques such as low-density parity-check (LDPC)~\cite{Liu2022LDPC} codes to add noise resistance. At the receiver end, when the SNR of the channel is sufficiently high, the decoder can perfectly recover the data. However, this approach performs poorly under low SNR conditions. To address this issue, the DeepJSCC framework was introduced. It jointly executes the encoding and decoding processes of source-channel communication through an end-to-end trained encoder-decoder pair (EDP), incorporating both the trainable and non-trainable components of the physical channel. This integrated approach optimizes the entire wireless transmission process from the original image to its reconstruction, effectively addressing the challenges caused by varying channel conditions.

Research has demonstrated that DeepJSCC outperforms traditional methods, such as JPEG/JPEG2000, especially under low SNR conditions~\cite{Zhang2022DeepJSCC}, offering superior data compression efficiency and enhanced noise-resistant transmission reliability. These advantages make DeepJSCC a highly effective solution for image transmission in challenging low SNR channel environments~\cite{Bourtsoulatze2019, Hu2022}.

Building on this, Xu {\it et al.}~\cite{Xu2021ADJSCC} proposed an attention-based deep joint source-channel coding (ADJSCC) model, which dynamically adjusts the learned image features based on the SNR of different channels. Inspired by methods that optimize compression ratios and channel coding rates, ADJSCC incorporates squeeze-and-excitation modules to adaptively adjust the image features in response to variations in SNR, thereby maximizing image reconstruction quality. Specifically, in addition to traditional convolutional and transposed convolutional layers, the model introduces Attention Feature Blocks (AFB)~\cite{Zhang2023AFB}. AFBs dynamically adjust the signal strength of the learned features, allowing the encoder and decoder to allocate resources between source coding/decoding and channel coding/decoding in a flexible manner. Simulation results show that, among multiple models trained under different SNR conditions, a single ADJSCC model achieves the highest peak signal-to-noise ratio (PSNR) performance, effectively addressing the challenge of SNR adaptation.

To further enhance the reliability of wireless image transmission, Yang {\it et al.}~\cite{Yang2022DeepJSCC-A} proposed a DeepJSCC method with adaptive rate control (DeepJSCC-A). This approach utilizes two separate encoders to handle source coding and channel coding independently, dividing the output semantic codes into selective and non-selective features. Selective features can be activated or deactivated based on specific requirements, while non-selective features remain consistently active.

In traditional communication systems, transmitters typically adjust the data transmission rate according to real-time channel conditions. Specifically, when using LDPC for channel coding, reducing the transmission rate can improve noise resistance under low SNR conditions. Additionally, employing lower modulation orders can further enhance transmission reliability. Traditional rate control methods rely on the relationship between channel SNR and transmission performance metrics, such as bit error rate (BER) and block error rate (BLER), which are typically predetermined.

The simulation results show that DeepJSCC-A can achieve adaptive rate control based on channel SNR and image content. However, it lacks the ability to adaptively adjust the compression rate (CR) when selecting image features, and its rate selection remains relatively limited. This means that, while DeepJSCC-A can adapt to varying SNR conditions, the model can only achieve optimal performance in wireless image transmission tasks when the channel environment closely matches the training conditions.

Unlike the rate selection framework of DeepJSCC-A, Zhang {\it et al.}~\cite{Zhang2023AFB} proposed a flexible and efficient variable code length DeepJSCC model, namely DeepJSCC-V. This model allows for arbitrary rate control based on different channel SNR and image content. Specifically, DeepJSCC-V uses a masking operation and a given CR to select the top $R$ proportion of semantic features as the transmission code, while discarding the remaining features. As a result, DeepJSCC-V can randomly generate CR values for different training samples without requiring additional training techniques. Compared to the DeepJSCC-A model, DeepJSCC-V achieves similar PSNR performance while providing higher transmission efficiency. It also employs an Oracle Network to optimize CR, enabling the model to adjust code lengths based on the channel environment.

However, as an emerging semantic communication solution, DeepJSCC still faces two key challenges: complex temporal channel denoising~\cite{Wu2024CDDM} and adaptation to dynamic channel variations~\cite{Ho2020Denoising}. While the DeepJSCC model offers significant flexibility in adjusting the bit rate, achieving effective information transmission in noisy channels remains a major challenge. The quality of signal transmission is influenced by multiple factors~\cite{Yang2022mulfactors}, including SNR, signal-to-interference-plus-noise ratio (SINR), and data content. The complex nonlinear relationships among these factors, as well as their impact on transmission quality, make effective channel modeling and representation highly challenging. Therefore, in addition to addressing the adaptation of channel SNR and CR, ensuring high-quality data transmission in noisy channel environments is equally critical.

The successful application of diffusion models (DM)~\cite{Croitoru2023DM} in the field of artificial intelligence-generated content (AIGC)~\cite{Liu2024AIGC} in recent years provides a novel approach to addressing the aforementioned challenges. During the forward diffusion process, the model continuously adds Gaussian noise to the original data until it becomes pure noise. Subsequently, in the reverse sampling process, the model is gradually trained to recover the original data from the noise. This process is similar to the recovery of clean features from a semantic signal transmitted through a noisy channel~\cite{Wu2024CDDM}. Inspired by this, we propose ADDM in this paper, which simulates noise and performs denoising based on different channel state information. Furthermore, we combine DeepJSCC with ADDM, and by simulating various channel environments, we significantly improve the performance of the semantic communication system in terms of both PSNR performance metric and visual quality.

\section{System Model}
In this section, we provide a detailed description of the system model for the adaptive semantic communication framework of CLEAR, focusing on the encoder-decoder architecture and the crucial role of CSI in parameter estimation and dynamic adaptation.

\vspace{-2pt}
\subsection{Encoder-Decoder Architecture}
CLEAR utilizes a trainable encoder-decoder architecture, which has been optimized to transmit and accurately reconstruct high-fidelity semantic information over complex time-varying channels. As shown in Fig.~\ref{fig:sys_model}, the architecture consists of four main components: the encoder, the wireless channel, the ADDM, and the decoder.

\subsubsection{Encoder}
The semantic encoder transforms the original data $S$ into a complex semantic code $f_{\theta_{e}}$. In the CLEAR system, the encoder consists of convolutional layers, batch normalization, residual connections, self-attention mechanisms, and adaptive feature blocks. 

First, the convolutional layer extracts low-level features using 64 $3\times3$ filters, while batch normalization and the ReLU activation function apply non-linear transformations to stabilize the data distribution. Then, residual connections and self-attention mechanisms capture global information, enhance long-range dependencies, and reduce spatial dimensions to $16\times16\times64$ via max pooling. Subsequently, adaptive feature blocks adjust dynamically to varying SNR values, optimizing feature representations and significantly improving adaptability. Through three successive encoding processes, the number of convolutional filters doubles at each layer, efficiently extracting semantic features $X_0$ from input data $S$, which are then transmitted over the wireless noisy channel. Specifically, the encoder is denoted by:
\begin{equation}
    X_0 = f_{\theta_{e}}(S, \theta_{e}),
\end{equation}
where $f_{\theta_{e}}$ represents the encoder function parameterized by $\theta_{e}$, and $X_0$ is the semantic feature output of the encoder.

\subsubsection{Wireless Channel}
The encoded semantic features $X_0$ are transmitted through a noisy wireless channel with complex, time-varying conditions, introducing distortions such as multipath effects, frequency-selective fading, phase noise, and Doppler shifts. CSI is used to estimate and adapt to these variations. The output signal $y_r$ from the noisy channel is given by:
\begin{equation}
    y_r = h \star X_0 + n,
\end{equation}
where $h$ represents the time-varying impulse response of the channel, $\star$ denotes the convolution operation, and $n$ is the additive Gaussian-distributed noise.

\begin{algorithm}[t]
\caption{Adaptive parameters updating for complex temporal-varying channels}
\begin{algorithmic}[1] 
\Require CSI
\Ensure Updated parameters
\State Initialize the parameters of complex temporal channels:
\[
h_{\text{init}} = 0, \; H_{\text{init}} = 0, \; \phi_{\text{init}} = 0, \; f_{\text{dinit}} = 0
\]
\State Estimate complex temporal channel parameters:
\[
h_{\text{est}} = [h_1(t), h_2(t), \ldots, h_L(t)]
\]
\[
H(f, t) = \int_{-\infty}^{\infty} h(t, \tau) e^{-j 2 \pi f \tau} d\tau
\]
\[
\phi_{\text{est}} = \phi(t), \quad f_d = \frac{v}{c} f_c \cos(\theta)
\]
\For{$i = 1$ \textbf{to} $L$}
    \State Compute $\alpha_i(t)$ and $\tau_i(t)$
    \State Update $h_i(t)$, $f_{d,i}$ and $\phi_i(t)$
\EndFor
\State Combine estimated parameters:
\[
\text{params}_{\text{updated}} = \sum_{i=1}^L \alpha_i(t) \delta(\tau - \tau_i(t)) e^{j (2\pi f_{d,i} t + \phi_i(t))}
\]
\State Apply updated parameters to the system
\State \Return Updated parameters
\end{algorithmic}
\end{algorithm}

\subsubsection{ADDM}
The obtained semantic features $X_0$ are input into the ADDM, where noise is gradually added to simulate a noisy channel. Using CSI, the noisy channel output $y_r$ is approximated, and progressive optimization
reduces channel noise to reconstruct the signal $\tilde{X}$ while minimizing distortions. The reverse denoising process of ADDM is expressed as:
\begin{equation}
    \tilde{X} = ADDM(X_0, y_r, \theta_{ADDM}).
\end{equation}
The detailed implementation process is provided in Section IV. 

\subsubsection{Decoder}
The semantic decoder comprises deconvolution layers, batch normalization, residual connections, self-attention mechanisms, and AFB. Unlike the feature extraction process of the encoder, the decoder starts with a deconvolution layer using 128 $3\times3$ filters to expand the size of the semantic feature map to $8\times8\times128$. Then, through a series of feature optimization processes similar to those in the encoder, the decoder transforms low-resolution semantic features into high-resolution outputs, effectively reconstructing the original image $\tilde{S}$ from the semantic features $\tilde{X}$ output of the ADDM. The decoding process is given by
\begin{equation}
    \tilde{S} = g_{\theta_{d}}(\tilde{X},\theta_{d}),
\end{equation}
where $g_{\theta_{d}}$ represents the decoder function parameterized by $\theta_{d}$, and $\tilde{S}$ is the reconstructed image.

\subsection{Role of CSI}
CSI is critical for enabling dynamical adaption to varying channel conditions in the CLEAR system. CSI provides essential information about the channel impulse response, including multipath propagation, fading coefficients, phase noise, and Doppler shifts. By utilizing CSI. By utilizing CSI, CLEAR estimates and adapts to real-time channel variations, ensuring robust performance across diverse channel environments.

\subsubsection{CSI Acquisition}
CSI is typically obtained through training sequences or pilot signals that are periodically transmitted with the data. These sequences allow the receiver to estimate the channel impulse response. The estimated CSI is then fed back to the transmitter to facilitate real-time adaptation.

\subsubsection{Adaptive Parameter Update}
The adaptive parameter updating process dynamically adjusts the noise parameters of the encoder and ADDM based on the estimated CSI. This process ensures that the system effectively mitigates the adverse effects of the channel, maintains high semantic fidelity, and minimizes distortion. The adaptive parameter updating process is summarized in Algorithm 1.

The algorithm first initializes all noise parameters to zero, establishing a baseline for the estimation and updating process. Next, the current channel conditions are estimated by processing the CSI and utilizing dedicated functions, including multipath effects, frequency-selective fading, phase noise, and Doppler shifts. The estimated parameters are then combined into a unified structure, providing a comprehensive view of the channel state. Using the unified structure, the system updates configurations, including filter settings, equalizer adjustments, phase noise compensation, and Doppler shift corrections.

\subsubsection{Multipath Effect}
Multipath effects~\cite{Manabe1996multipath} arise when a signal arrives at the receiver via multiple paths, each with distinct delays and attenuation, which is typically represented by a time-varying impulse response, as
\begin{equation}
    h(t, \tau) = \sum_{i=1}^L \alpha_i(t) \delta(\tau - \tau_i(t)),
\end{equation}
where $\alpha_{i}(t)$ is the time-varying fading coefficient for the $i$-th path, $\tau_i(t)$ is the time-varying delay for the $i$-th path, $L$ is the number of multiple paths, and $\delta$ is the Dirac delta function.

\begin{algorithm}[t]
\caption{Diffusion algorithm for ADDM}
\begin{algorithmic}[1]  
\Require Initial signal $x_0$, channel impulse response $h(t,\tau)$, noise parameter sequence $\{a_t\}_{t=0}^T$, time steps $T$
\Ensure Noisy signal sequence
\State Initialize parameters $x_0 \leftarrow X$
\For{$t = 1$ \textbf{to} $T$}
    \State Generate standard normal noise: $\epsilon_t \sim N(0, I)$
    \State Compute the mixed signal and noise:
    \[
    x_t = \sqrt{\alpha_t} \int_{-\infty}^\infty h(t, \tau) x_0(t - \tau) d\tau + \sqrt{1 - \alpha_t} \epsilon_t
    \]
    \State Store $x_t$ in the sequence
\EndFor
\State \Return the noisy signal sequence
\end{algorithmic}
\end{algorithm}

\subsubsection{Frequency-Selective Fading}
Frequency-selective fading~\cite{Liu2002frequency-selective-fading} is characterized in the frequency domain by the channel transfer function, derived from the Fourier transform of the time-domain channel impulse response, as
\begin{equation}
    H(f, t) = \int_{-\infty}^{\infty} h(t, \tau) e^{-j 2 \pi f \tau} d\tau.
\end{equation}

\subsubsection{Phase Noise}
Phase noise~\cite{Robins1984Pn} caused by oscillator imperfections, introduces random phase disturbance. It is expressed as
\begin{equation}
    X(t) = X(t) e^{j \phi(t)},
\end{equation}
where $X(t)$ is the input signal, and $\phi(t)$ is the phase noise.

\subsubsection{Doppler Effect}
The Doppler effect~\cite{Blaugrund1966Ds}, resulting from relative motion between the transmitter and receiver, induces frequency shifts of the signal. It is given by
\begin{equation}
    f_d = \frac{v}{c}f_c \cos\theta,
\end{equation}
where $f_d$ is the Doppler shift, $v$ is the relative velocity, $c$ is the speed of light, $f_c$ is the carrier frequency, and $\theta$ is the angle between the direction of movement and the signal propagation direction. The time-varying impulse response with Doppler shifts is given by:
\begin{equation}
    h(t, \tau) = \sum_{i=1}^L \alpha_i \delta(\tau - \tau_i) e^{j 2 \pi f_{d,i} t}.
\end{equation}

Combining multipath effects, frequency-selective fading, phase noise, and Doppler shift, the time-varying channel model is:
\begin{equation}
    h(t, \tau) = \sum_{i=1}^L \alpha_i(t) \delta(\tau - \tau_i(t)) e^{j (2 \pi f_{d,i} t + \phi(t))}.
\end{equation}

By extracting CSI, parameters $\alpha_i(t)$, $\tau_i(t)$, $f_{d,i}$ and $\phi(t)$ can be estimated to reconstruct the channel environment. With a large number of multiple paths $L$, the central limit theorem ensures that the whole channel response approximates a complex Gaussian distribution.

The adaptive parameter update dynamically adjusts noise parameters based on estimated CSI values. Initially, system parameters are configured according to the number of paths. Specialized functions process the CSI to estimate parameters for multipath effects, fading, phase noise, and Doppler shifts. These parameters are integrated and used to update $h(t, \tau)$, enabling the system to adapt to channel conditions. This process also optimizes the ADDM, ensuring CLEAR responds effectively to varying environments.

\section{Adaptive Diffusion Denoising Model}
The ADDM is a core component of the CLEAR system, designed to simulate noisy channel outputs and progressively remove channel noise. This process enhances signal quality, minimizes distortion, and ensures high semantic fidelity. This section details the structure of ADDM, noise diffusion training, and sampling algorithms.

\begin{algorithm}[t]
\caption{Sampling algorithm of ADDM}
\begin{algorithmic}[1]
\Require Noisy signal sequence, channel impulse response $h(t,\tau)$, noise parameter sequence $\{a_t\}_{t=0}^T$, denoising mean network $\mu_\theta$, denoising variance network $\sigma_t$, time steps $T$
\Ensure Restored signal
\State Initialize parameters $x_0 \leftarrow X_T$
\For{$t = T$ \textbf{to} $1$}
    \State Compute the denoised mean: 
    \[
    \mu_t(x_t) = \frac{1}{\sqrt{\alpha_t}} \left( x_t - \sqrt{1 - \alpha_t} \epsilon_t \right)\]
    \State Generate standard normal noise: $\epsilon_t' \sim N(0, I)$
    \State Update the signal: $x_{t-1} = \mu_t(x_t) + \sigma_t \epsilon_t'$
    \State Consider channel influence: \[ x_{t-1} = \int_{-\infty}^\infty h(t, \tau) x_{t-1}(t - \tau) d\tau + \sigma_t \epsilon_t'\]
\EndFor
\State \Return the restored signal
\end{algorithmic}
\end{algorithm}

\subsection{ADDM Structure and Functions}
The ADDM reduces noise in complex, time-varying wireless channels using a specialized noise scheduling mechanism. Its structure consists of two main stages: the forward diffusion process and the reverse denoising process.

As shown in Fig.~\ref{fig:sys_model}, ADDM leverages an improved U-Net architecture to recover image semantic features degraded by channel distortions. The encoder compresses input data into a latent space via downsampling, preserving core semantic features while reducing dimensionality for efficient transmission. The decoder reconstructs the image through gradual upsampling, supported by skip connections that transfer intermediate encoder features directly to corresponding decoder layers. These connections preserve critical details and spatial information, enhancing the recovery of clean semantic features during denoising.

To further improve feature reconstruction, ADDM incorporates residual convolution layers and self-attention mechanisms at each upsampling step. These components enhance feature extraction and enable deeper learning of image semantics. Additionally, ADDM processes the real and imaginary components of complex-valued data separately, making it particularly effective at addressing time-varying channel effects, such as Doppler shifts and phase noise. This specialized design ensures robust performance even in challenging communication scenarios.

From a functional perspective, ADDM offers a robust, adaptive solution for semantic communication in complex channels. By training on various noise types, ADDM enhances the adaptability of the CLEAR system, effectively handling diverse channel distortions while keeping semantic integrity. Its improved U-Net architecture excels in retaining detailed image features during denoising, ensuring accurate data reconstruction. This design not only stabilizes the semantic communication system but also highlights the potential of advanced deep-learning techniques in real-world communication scenarios.
\begin{algorithm}[t]
\caption{Training algorithm of CLEAR system}
\begin{algorithmic}[1]
\Require Training data $S$, noise parameter sequence $\epsilon_t$, time steps $T$, CSI
\Ensure Trained parameters for CLEAR system
\State Initialize parameters $\theta_e$, $\theta_d$, $\theta_{ADDM}$
\While {semantic loss is not converged}
    \State Encode the data: $X_0 = f_{\theta_e}(S; \theta_e)$
    \State Simulate the forward diffusion process to generate noisy signal sequence ${X_t}$
    \State Apply ADDM to diffuse noise, then denoise the signal:
    \[
    \tilde{X} = ADDM(X_0, y_r, \theta_{ADDM})\]
    \State Decode the denoised signal: $\tilde{S} = g_{\theta_d}(\tilde{X}; \theta_d)$
    \State Compute the end-to-end loss: $L = Loss(\tilde{S},S)$
    \State Update parameters: $\theta_e$, $\theta_d$, $\theta_{ADDM}$ using gradient descent
\EndWhile
\State \Return trained parameters
\end{algorithmic}
\end{algorithm}

\subsubsection{Forward Diffusion Process}
In ADDM, the forward process gradually adds noise to the data, simulating a noise channel. The initial semantic features are denoted as $x_{0}$ with the noise represented by $\epsilon$. The forward process is then expressed as:
\begin{equation}
    x_t = \sqrt{\alpha_{t}}x_{t-1} + \sqrt{1 - \alpha_{t}}\epsilon,
\end{equation}
where $x_t \in x_0$, $\alpha_{t}$ is a noise parameter which decreases over time, and $\epsilon$ follows a standard normal distribution. 

In time-varying wireless channels, the received signal is treated as the result of a noisy transmission, and thereby the forward process is expressed as
\begin{equation}
    x_t = \int_{-\infty}^{\infty} h(t, \tau) x_0(t - \tau) d\tau + \sqrt{1 - \alpha_t} \epsilon.
    \label{eq:foward_diffusion}
\end{equation}

When $\alpha_t \to \alpha_T$ and $\alpha_T \to 0$, we obtain
\begin{equation}
    x_T = \sqrt{0} \int_{-\infty}^{\infty} h(T, \tau) x_0(T - \tau) d\tau + \sqrt{1} \epsilon_T = \epsilon_T,
\end{equation}
indicating that $x_T$ is pure noise, where the proof is summarized in Appendix A. 

\subsubsection{Reverse Denoising Process}
The reverse process gradually removes noise to reconstruct the original signal, which is expressed as:
\begin{equation}
    p(x_{t-1} | x_t) = N(x_{t-1}; \mu(x_t), \sigma_t^2 I),
\end{equation}
where $\mu(x_t)$ and $\sigma_t$ represent the estimated denoising mean and variance, respectively. 

Therefore, the entire process can be viewed as the inverse operation of the forward diffusion process in \eqref{eq:foward_diffusion}, as
\begin{equation}
    x_{t-1} = \frac{1}{\sqrt{\alpha_t}} ( x_t - \int_{-\infty}^{\infty} h(t, \tau) x_{t-1}(t - \tau) d\tau) + \sqrt{1 - \alpha_t} \epsilon_t.
\end{equation}

By gradually removing noise, the reverse process ultimately restores the original signal, as
\begin{equation}
    x_0 = \frac{\lim_{t \to 0} \left( x_t - \sqrt{1 - \alpha_t} \epsilon_t\right)}{\sqrt{\alpha_t}}.
\end{equation}
A detailed proof of this process is given in Appendix B.

\begin{figure}[t]
    \centering
     \includegraphics[width=0.4\textwidth]{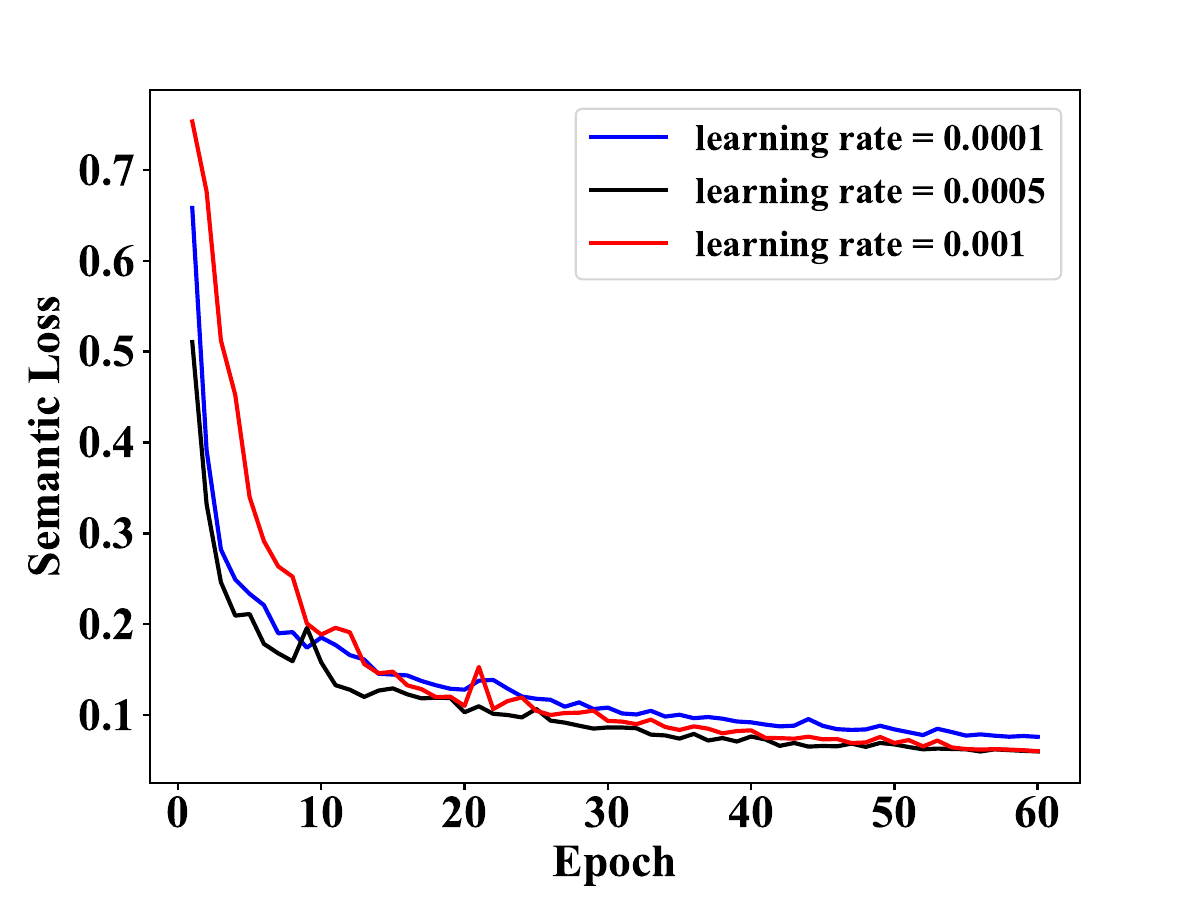}
    \caption{The training loss in terms of epochs in the CLEAR system.}
    \label{fig:loss}
\end{figure}
\begin{figure*}[t]
 \centering
   \subfigure[CIFAR10 dataset]{
		\begin{minipage}[t]{0.4\linewidth}
            \centering
            \includegraphics[width=0.96\textwidth]{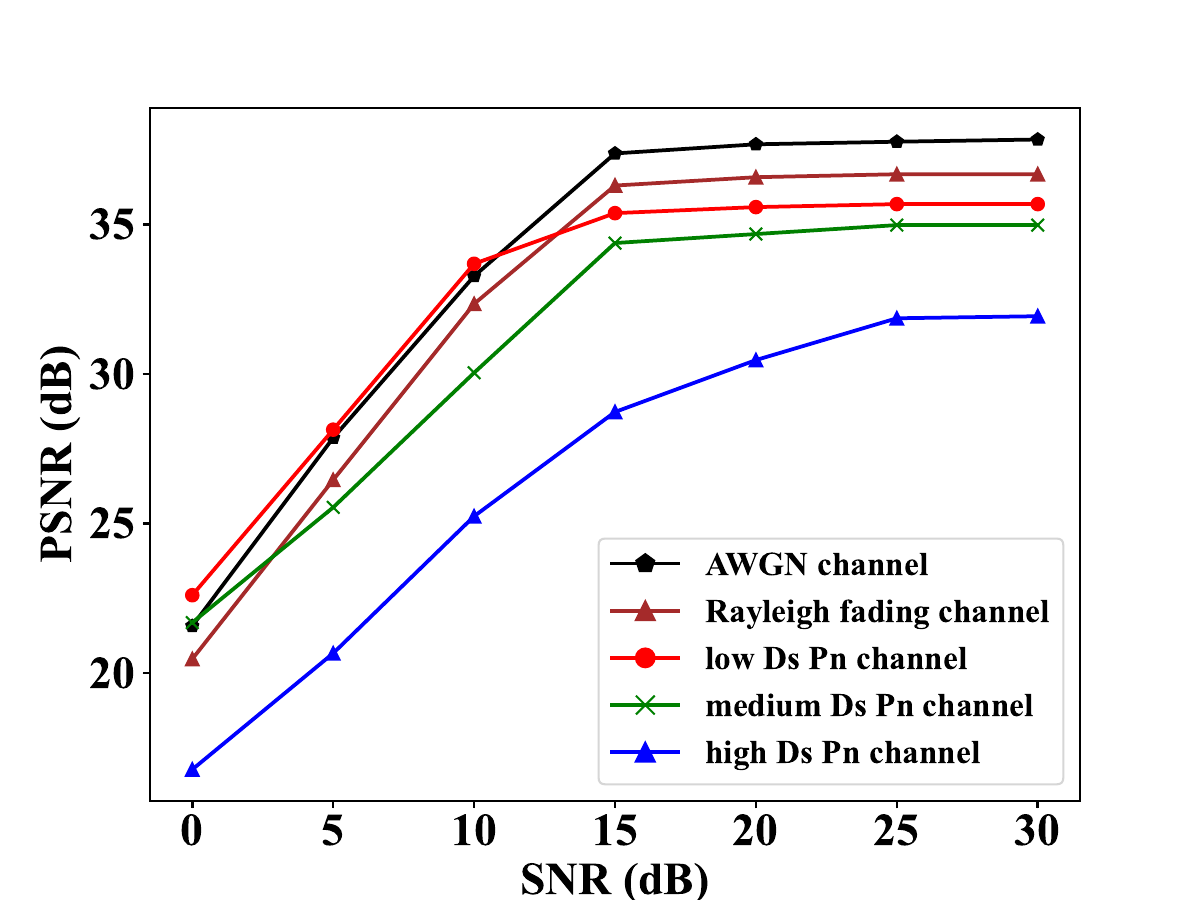}
            \label{fig:reslut_cifar10}
            \end{minipage}
		}
    ~
    \subfigure[DIV2K dataset]{
		\begin{minipage}[t]{0.4\linewidth}
            \centering
            \includegraphics[width=0.96\textwidth]{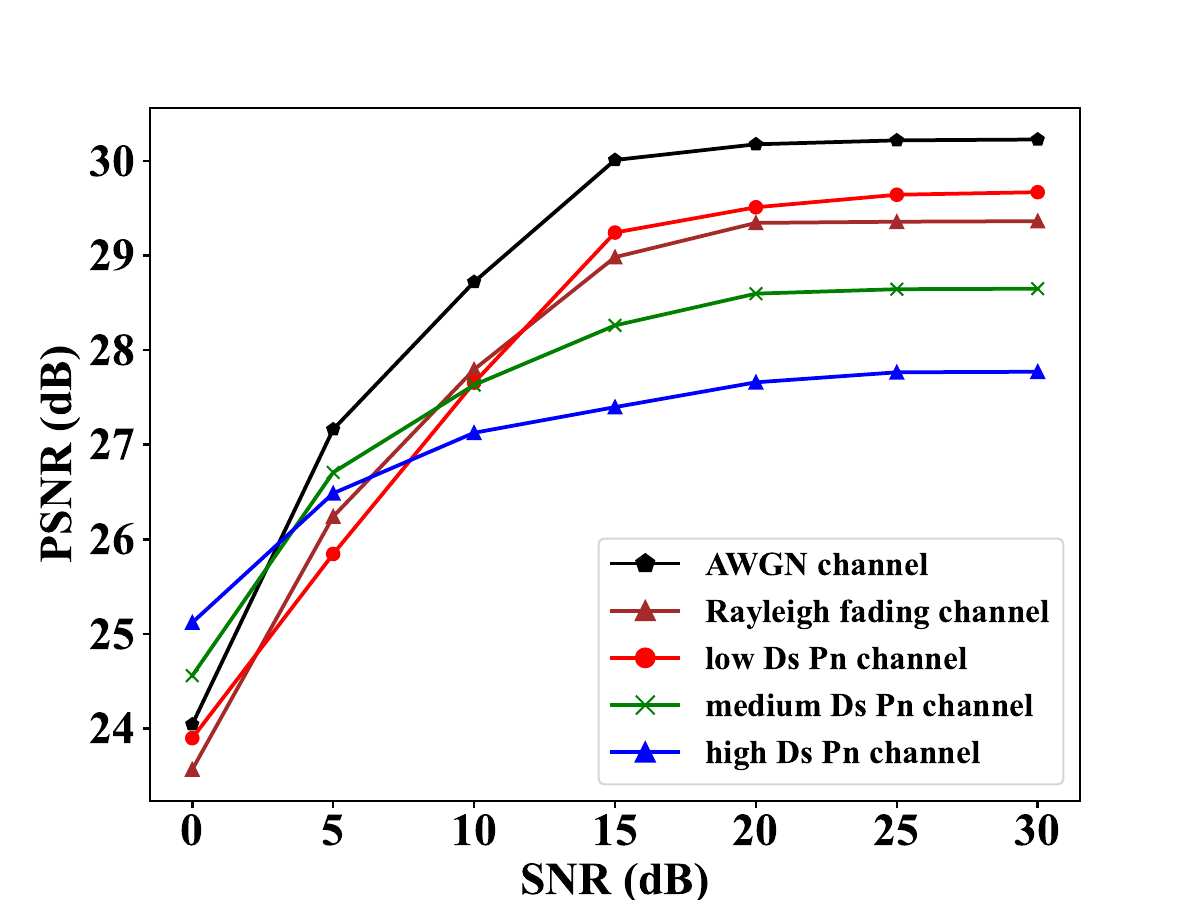}
            \label{fig:reslut_div2k}
            \end{minipage}
		}
\caption{PSNR performance on CIFAR10 and DIV2K datasets under various channel conditions and SNR values. $Ds$ stands for Doppler shifts, and $Pn$ represents phase noise.}
\label{fig:results_on_AWGN_and_DIV2K}
\end{figure*}

\subsection{Diffusion Algorithm of ADDM}
The forward diffusion algorithm of ADDM trains the model to effectively add diffusion noise and perform denoising sampling on the received signals. The core objective of the training process is to learn the noise distribution and minimize the loss between the predicted noise and the actual noise, ensuring robust denoising capabilities.

The entire training process ensures that ADDM can accurately simulate the noise characteristics of the wireless channel, enabling effective denoising during the sampling phase. The detailed steps of the training process are outlined in Algorithm 2. 

First, samples are extracted from the data, and a random time $t$ is selected from $1$ to $T$. Then, $x_0$ and $t$ are passed to the Gaussian diffusion model, where random noise is sampled and added to $x_0$, resulting in $x_t$, as defined by the forward diffusion process. After that, $x_t$ and $t$ are input into the ADDM, which adopts an improved U-Net architecture. ADDM combines $t$ and $x_t$ to generate a sine position encoding, predicts the simulated noise, and returns the result. The Gaussian diffusion calculates the loss between the predicted noise and the random noise. Also, it computes the L2 loss between the predicted noise and the previously sampled random noise from the Gaussian diffusion. Finally, gradient calculation and weight updates are performed to optimize the model. These steps are repeated for multiple iterations until the ADDM network is fully trained. This iterative training process ensures that ADDM can accurately learn the noise distribution of the wireless channel and perform precise denoising during the sampling phase.

\subsection{Sampling Algorithm of ADDM}
In the denoising phase, the ADDM employs a progressive sampling algorithm to reconstruct the original signal from the noisy received signal. This algorithm iteratively removes noise from the received signal using the trained ADDM parameters, enabling accurate reconstruction the original semantic features. 

The training process is outlined in Algorithm 3. It repeats the following steps: First, noise is sampled from a standard normal distribution and performed reparameterization; Then, the trained ADDM computes the mean of the noise using the noise input signal; After that, the denoised result $x_t$ is combined with the noise variance $\sigma_t$; Finally, parameterization is applied to obtain $x_{t-1}$ and generates the reconstructed signal.

\subsection{Training Algorithm of CLEAR}
The CLEAR system integrates ADDM and DeepJSCC, combining their strengths to enable efficient and reliable semantic communication in complex, time-varying channel environments. The training process jointly optimizes the semantic encoder-decoder and ADDM, minimizing end-to-end loss and ensuring robust performance under various channel conditions.

The training of the CLEAR system begins with initializing the parameters $\theta_e$, $\theta_d$, and $\theta_{ADDM}$, which correspond to the semantic encoder, decoder, and ADDM, respectively. The encoder $f_{\theta_e}$ compresses the input data $S$, extracting semantic features $X_0$. Using Algorithms 1 and 2, ADDM simulates a noisy signal sequence $X_t$ by progressively adding noise through the forward diffusion process, incorporating real-time CSI to reflect complex channel conditions. The denoising phase then uses Algorithm 3 to iteratively remove the noise and recover clean semantic features $\tilde{X}$. Subsequently, the decoder $g_{\theta_d}$ reconstructs the denoised signal $\tilde{S}$ from $\tilde{X}$. 

To ensure the reconstructed signal closely matches the original input $S$, the mean squared error (MSE)~\cite{Sara2019MSE} is computed as the semantic loss $L$, measuring the similarity between $\tilde{S}$ and $S$. The parameters $\theta_e$, $\theta_d$, and $\theta_{ADDM}$ are then updated using the Adam optimizer, minimizing $L$ iteratively through backpropagation. This process continues until the model converges, producing an optimized encoder, decoder, and ADDM parameters. These trained components enable the CLEAR system to achieve high-quality semantic communication and denoising performance in complex, time-varying channel environments.

 \begin{figure*}[t]
 \centering
   \subfigure[compression rate 0.3]{
		\begin{minipage}[t]{0.4\linewidth}
            \centering
            \includegraphics[width=0.96\textwidth]{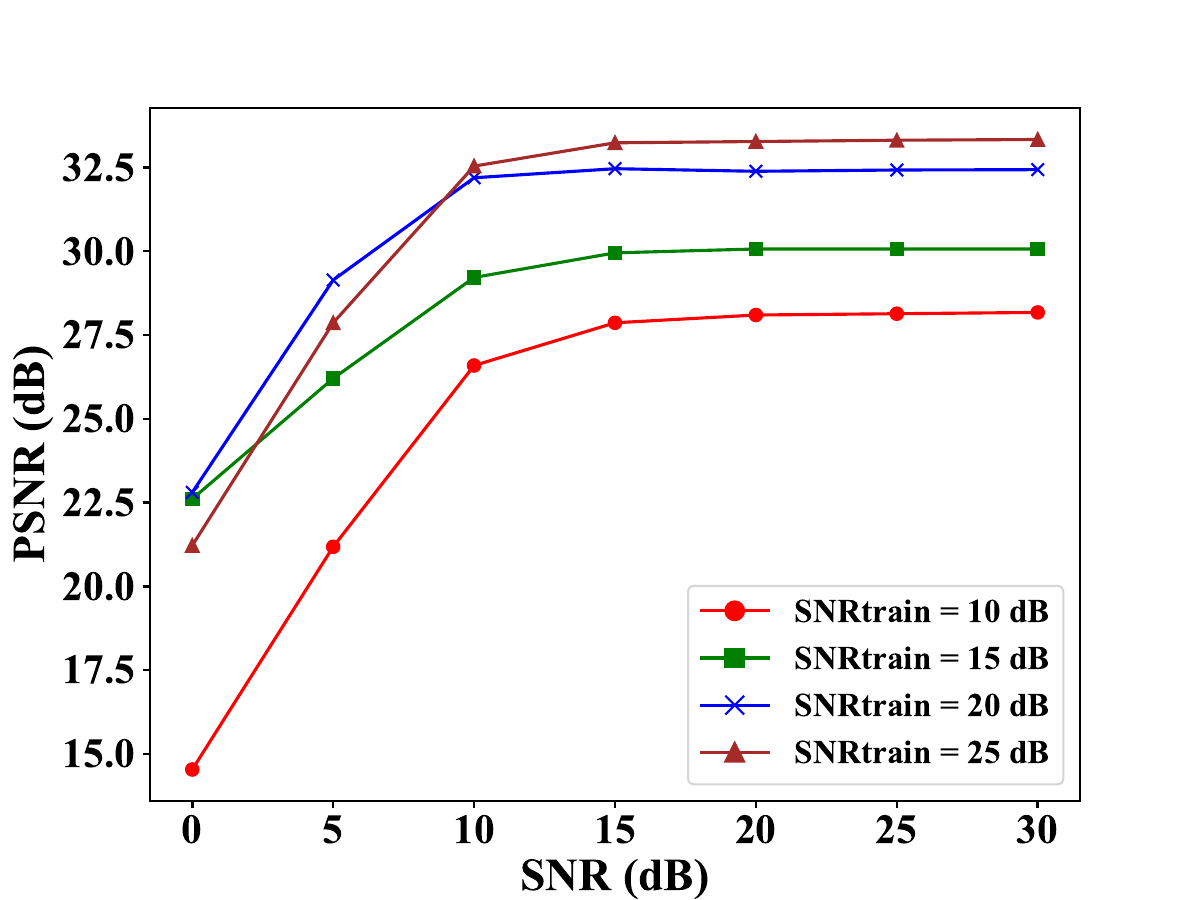}
            \label{fig:reslut_awgn_0_3}
            \end{minipage}
		}
    ~
    \subfigure[compression rate 0.6]{
		\begin{minipage}[t]{0.4\linewidth}
            \centering
            \includegraphics[width=0.96\textwidth]{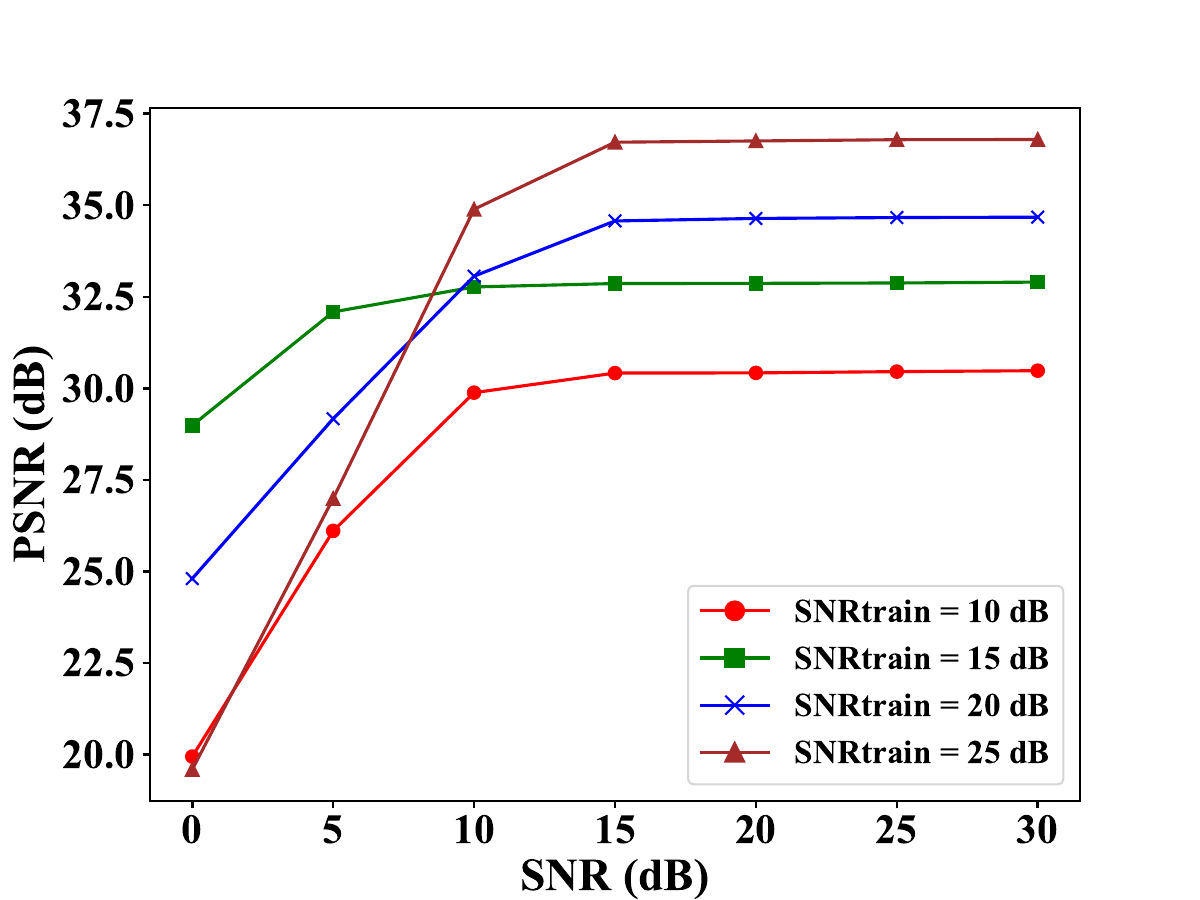}
            \label{fig:reslut_awgn_0_6}
            \end{minipage}
		}
\caption{Effect of compression rate on PSNR performance in AWGN channels with varying SNR values.}
\label{fig:results_on_AWGN}
\end{figure*}
\begin{table*}[t]
\centering
\caption{Datasets and simulation settings in the evaluations.}
\label{tab:datasets}
\renewcommand{\arraystretch}{1.6}
\begin{tabular}{lccc}
\hline
\textbf{Dataset}& \textbf{Resolution}  & \textbf{Usage}  & \textbf{Settings} \\ \hline
CIFAR-10  & $32 \times 32$ & Training: 50,000, Validation: 10,000 & Batch: 128,  Learning rate $10^{-4}$ \\ 
CIFAR-100 & $32 \times 32$ & Test: 10,000 & Evaluation with unseen data \\ 
DIV2K    & $\sim2048 \times 1080$ & Training: 800, Validation: 100, Test: 100 & Crop: $256 \times 256$, Batch: 32, Learning rate: $10^{-3}$ \\ 
Kodak24   & $768 \times 512$ & Test: 24   & 10 repeats for evaluating PSNR of JPEG, ADJSCC, DeepJSCC-V \\ \hline
\end{tabular}
\end{table*}

To provide a clearer illustration of the model's optimization process during training, Fig.~\ref{fig:loss} illustrates the variation in loss over 60 training epochs. We simulated typical real-world communication scenarios using moderate noise intensity (with Doppler shift set to 0.2, phase noise set to 0.2, and an SNR of 15~dB). It can be seen that when the learning rate is set to $10^{-4}$ the semantic loss converges the slowest and reaches the highest value. As training progresses, although there are some fluctuations in the semantic loss for all three learning rates during the decline, the loss values generally decrease and converge around the 50th epoch. These results highlight the CLEAR system’s ability to continuously optimize and improve performance during training, even in dynamic and noisy channel environments.

The following section presents detailed experimental results validating the effectiveness of the CLEAR system framework and demonstrating its superiority over existing methods in various challenging scenarios.

\section{Experimental Results}
This section describes the experimental setups and presents a series of experimental results, evaluating the robustness of the CLEAR system under various channel environments and SNR values. It demonstrates the effectiveness and adaptability of the proposed model.

\subsection{Datasets and Simulation Settings}
The performance of the CLEAR system is evaluated on three benchmark datasets: CIFAR-10, CIFAR-100 datasets~\cite{Krizhevsky2009CIFAR}, and DIV2K dataset~\cite{Agustsson2017DIV2K}. Additionally, the Kodak24 dataset~\cite{Kodak24} is used for visual comparisons with other image transmission schemes, where the details are shown in Table~\ref{tab:datasets}. The maximum epoch for the algorithm is set at 100 with an early stopping mechanism~\cite{Caruana2000earlystop}.

\subsection{Adaptive Network Denoising Experiment}
We conducted two representative image restoration experiments using CLEAR to evaluate its effectiveness. These experiments were performed on CIFAR-10 and DIV2K datasets. Leveraging a deep learning-based approach, we integrated a semantic encoder-decoder architecture with the ADDM to restore images and mitigate the effects of channel distortions. The experiments were carried out under various channel conditions, including additive white Gaussian noise (AWGN), Rayleigh fading, and complex time-varying channels characterized by different Doppler shifts and phase noise levels. The results demonstrate that CLEAR effectively preserves image quality across a wide range of SNR conditions, as measured by the PSNR metric.
\begin{figure*}[t]
 \centering
   \subfigure[compression rate 0.3]{
		\begin{minipage}[t]{0.4\linewidth}
            \centering
            \includegraphics[width=0.96\textwidth]{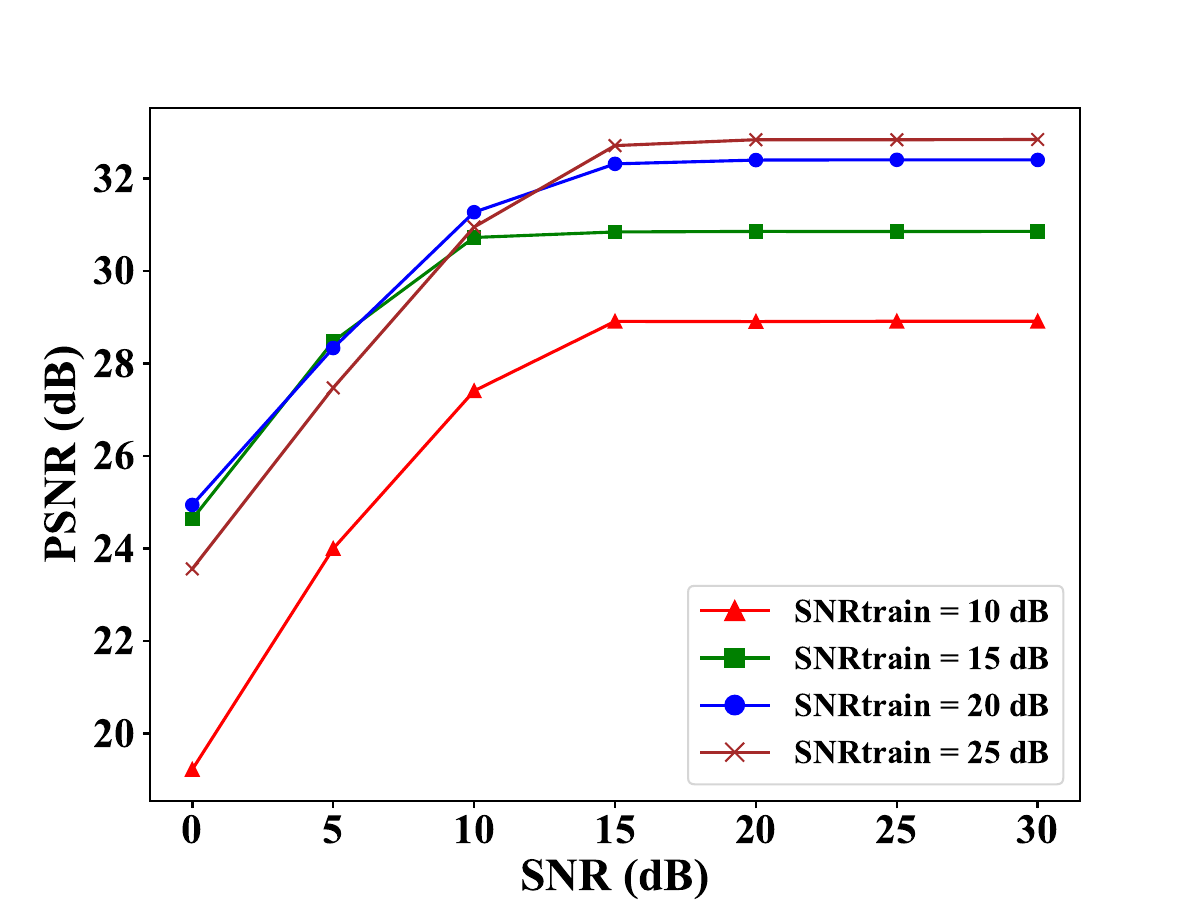}
            \label{fig:reslut_rayleigh_0_3}
            \end{minipage}
		}
    ~
    \subfigure[compression rate 0.6]{
		\begin{minipage}[t]{0.4\linewidth}
            \centering
            \includegraphics[width=0.96\textwidth]{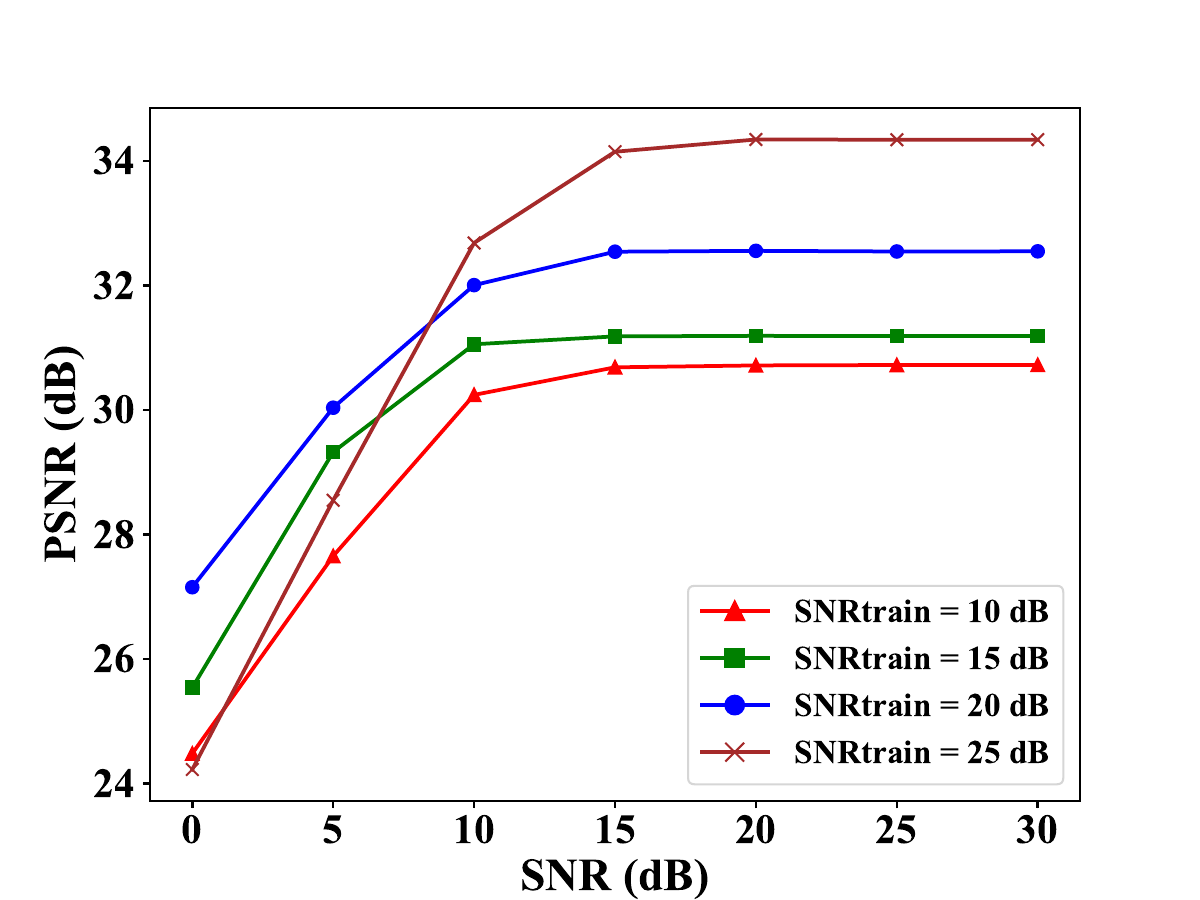}
            \label{fig:reslut_rayleigh_0_6}
            \end{minipage}
		}
\caption{Effect of compression rate on PSNR performance in Rayleigh fading channel at varying SNR.}
\label{fig:results_on_Rayleigh}
\end{figure*}

\subsubsection{Impact of datasets}
Fig.~\ref{fig:results_on_AWGN_and_DIV2K} shows the PSNR performance of the CLEAR system on the CIFAR-10 and DIV2K datasets under various channel conditions. The results highlight the effectiveness of CLEAR in restoring transmitted images and mitigating channel distortions.

For both datasets, PSNR improves steadily as the SNR increases across all channel conditions. The AWGN channel achieves the highest PSNR due to its relatively simple noise characteristics, while the Rayleigh fading channel exhibits slightly reduced performance due to multipath fading effects. In complex time-varying channels, the performance depends on the Doppler shift ($Ds$) and phase noise ($Pn$). We consider three environments with different $Ds$ of $0.01$, $0.05$, and $0.5$, representing low, medium, and high Doppler shifts, respectively. The corresponding $Pn$ values of $0.01$, $0.05$, and $0.2$ represent the best to worst cases. Higher Doppler shifts and phase noise levels lead to greater PSNR degradation, particularly for high-resolution images in the DIV2K dataset. However, CLEAR demonstrates resilience by maintaining high PSNR values even under challenging conditions, with performance stabilizing after an SNR of $20$~dB. This demonstrates the robustness in preserving image quality across varying resolutions of the CLEAR system.

\begin{figure*}[t]
 \centering
   \subfigure[compression rate 0.3]{
		\begin{minipage}[t]{0.4\linewidth}
            \centering
            \includegraphics[width=0.96\textwidth]{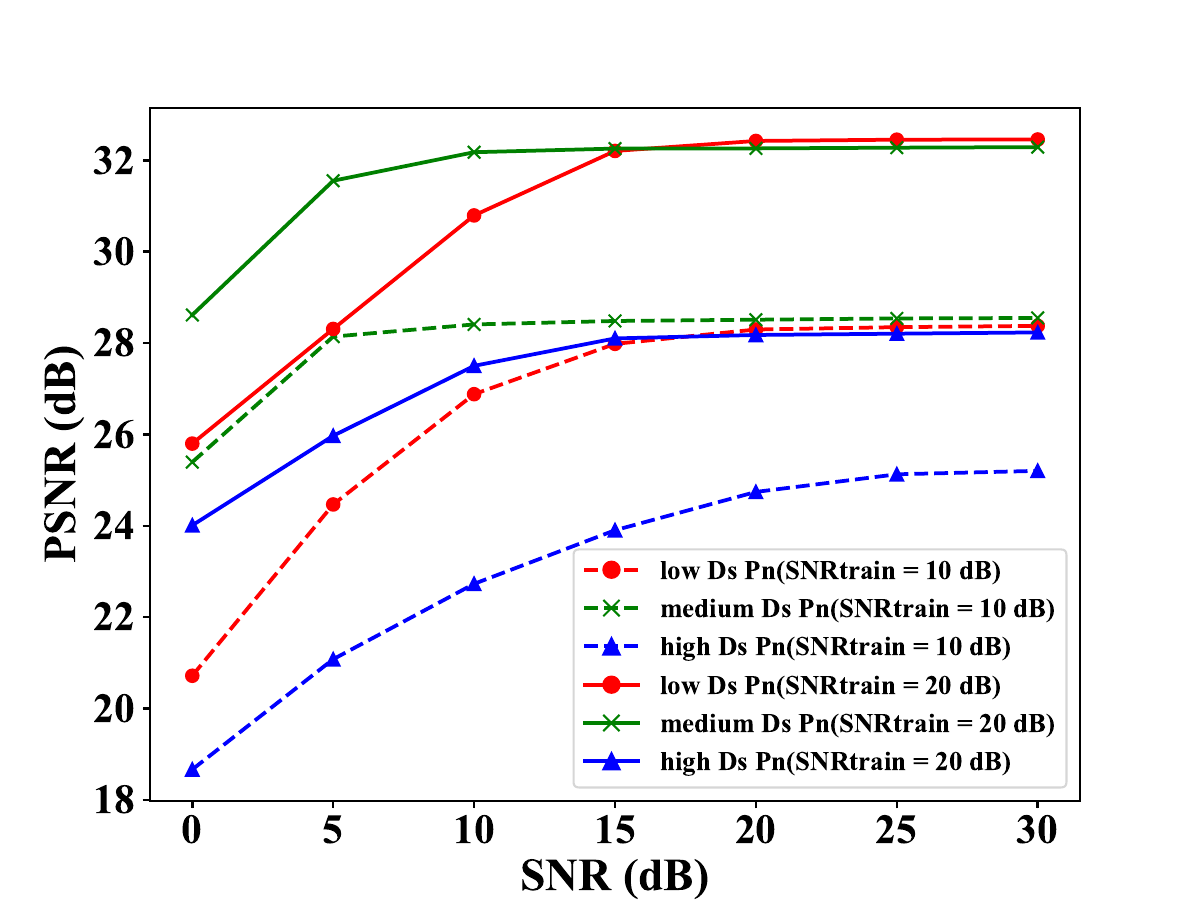}
            \label{fig:reslut_complex_0_3}
            \end{minipage}
		}
    ~
    \subfigure[compression rate 0.6]{
		\begin{minipage}[t]{0.4\linewidth}
            \centering
            \includegraphics[width=0.96\textwidth]{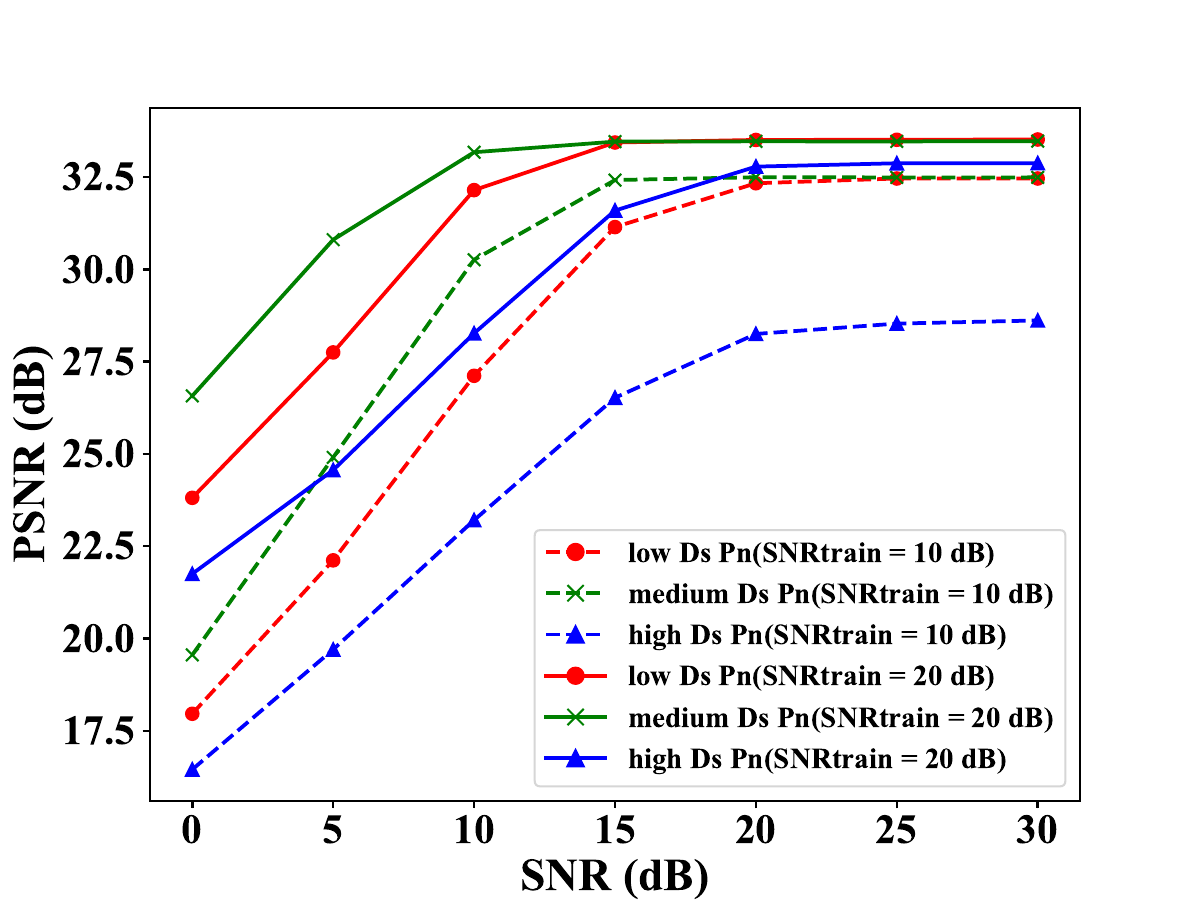}
            \label{fig:reslut_complex_0_6}
            \end{minipage}
		}
\caption{Effect of compression rate on PSNR performance in the complex time-varying channels with different Doppler effects and SNR values.}
\label{fig:results_on_Complex}
\end{figure*}

\subsubsection{Impact of different channels}
We first evaluate the robustness of the CLEAR system in the AWGN channel at different SNR levels and compression rates. Fig.~\ref{fig:results_on_AWGN} illustrates the PSNR variations for reconstructed images at two compression rates, $0.3$ and $0.6$, respectively. The results demonstrate that higher compression rates positively influence image quality recovery. All results are trained and optimized under specific channel SNR conditions, and the performance is evaluated on test images with varying SNR levels.

As the test SNR ($SNR_{test}$) increases, the quality of reconstructed images steadily improves. At a compression rate of $0.6$, all results corresponding to different $SNR_{train}$ levels reach saturation nearly simultaneously after $SNR_{test} = 15$~dB. In contrast, for a compression rate of $0.3$, the growth rates of the curves vary slightly, with saturation depending more on the $SNR_{train}$ setting. Notably, Fig.~\ref{fig:results_on_AWGN}(a) and (b) show that when $SNR_{train} \geq 15$~dB, higher $SNR_{train}$ values result in wider PSNR spans, characterized by lower initial PSNR values but higher peaks. Across all SNR levels, the test performance consistently outperforms the case where $SNR_{train} = 10$~dB, emphasizing the significant impact of $SNR_{train}$ on image reconstruction quality.

Fig.~\ref{fig:results_on_Rayleigh} evaluates the PSNR performance of the CLEAR system under Rayleigh fading channel conditions. Similar to the AWGN case, two compression rates, $0.3$ and $0.6$, are analyzed. Compared to the static AWGN channel, the overall performance declines slightly due to the additional complexity of fading effects. However, the reconstructed images remain of acceptable quality, demonstrating that CLEAR effectively adapts to varying channel conditions.

At a compression rate of $0.6$, the saturation point across different $SNR_{train}$ levels is consistent with the AWGN case, occurring after $SNR_{test} = 15$~dB. However, the rate of improvement is slower, reflecting the challenges posed by fading. At a compression rate of $0.3$, the curves show more pronounced variations, particularly under low $SNR_{test}$ conditions. These results highlight the model's ability to estimate and adapt to CSI, dynamically adjusting the ADDM parameters to mitigate fading effects.

By comparing Fig.~\ref{fig:results_on_AWGN} and \ref{fig:results_on_Rayleigh}, we observe that the PSNR trends for both channels are generally similar. However, in the Rayleigh fading channel, the performance shows a stronger dependence on $SNR_{train}$, resulting in less stability, particularly at low compression rates and $SNR_{test}$ values. Additionally, under Rayleigh fading, curves corresponding to higher $SNR_{train}$ levels exhibit slower growth, with a temporary recovery deviation observed at $SNR_{test} = 5$~dB. This deviation may be attributed to the system's limited adaptability under such conditions, delaying parameter adjustments. Despite these challenges, the CLEAR system demonstrates robust performance, maintaining acceptable reconstruction quality across both channels.

We further examine the robustness of CLEAR in complex time-varying channels, characterized by combinations of $D_s$ and $P_n$ at varying levels. Fig.~\ref{fig:results_on_Complex} presents the PSNR trends for two compression rates, \ie, $0.3$ and $0.6$. Under high compression rates and $SNR_{train}$ levels, the PSNR trends align closely with those observed in the AWGN and Rayleigh channels. As $SNR_{test}$ increases, the PSNR improves significantly, with reconstructed image quality nearing that of the Rayleigh fading channel at saturation.

The effects of Doppler shifts and phase noise become more pronounced in complex channels. Higher Doppler shifts and phase noise levels lead to greater PSNR degradation. However, Fig.~\ref{fig:results_on_Complex}(a) and (b) show that the CLEAR system maintains visually acceptable reconstruction quality even under high noise conditions, thanks to the ADDM module. Interestingly, at $SNR_{test} \leq 10$~dB, medium noise levels outperform low-noise channels in terms of reconstruction quality, suggesting complex interactions between Doppler shifts and phase noise. This nonlinear behavior requires further investigation.

\begin{figure}[t]
    \centering
     \includegraphics[width=0.4\textwidth]{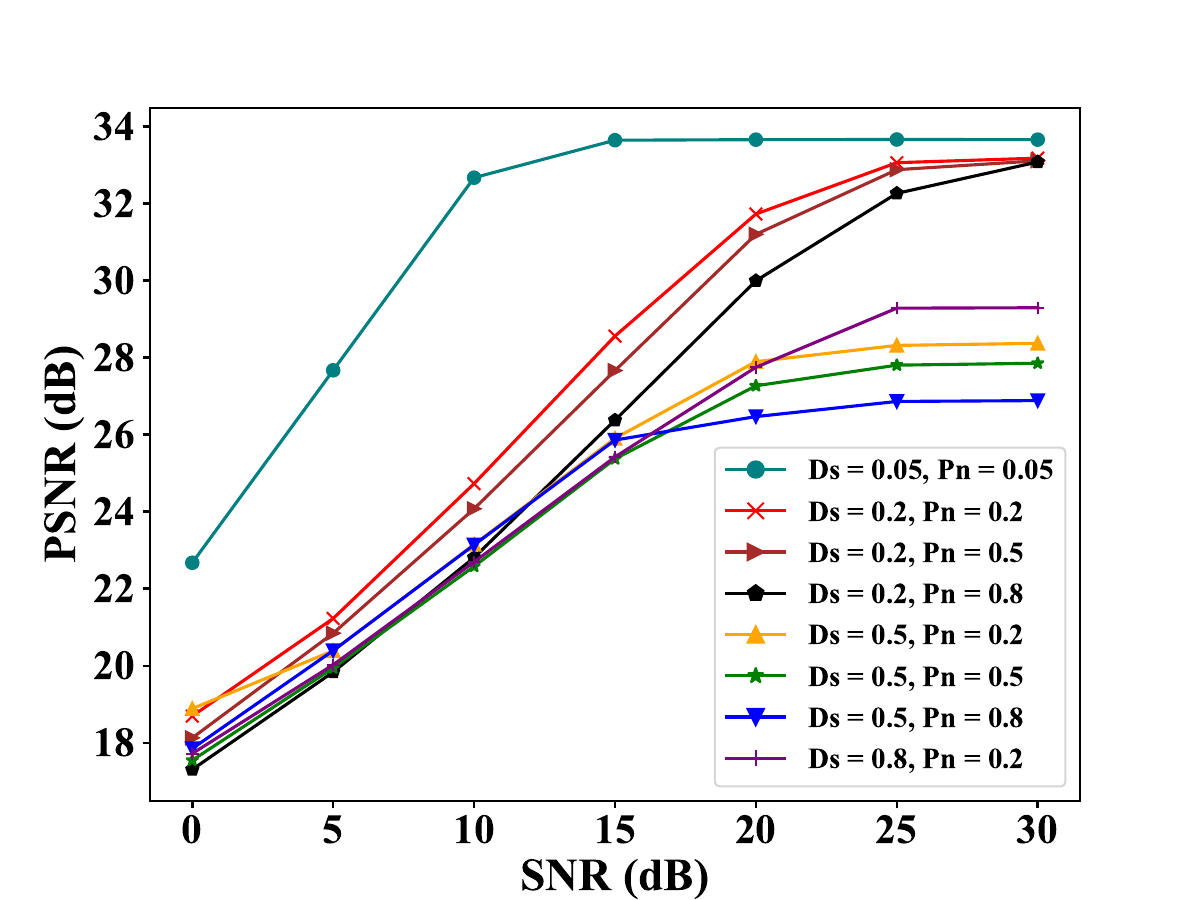}
    \caption{The comparison of PSNR performance under different Doppler shifts and phase noise combinations in time-varying channels.}
    \label{fig:Compare_Complex}
\end{figure}

\subsubsection{Impact of Doppler shifts and phase noise}
To analyze the interplay between Doppler shifts and phase noise, we studied various combinations of these parameters to simulate complex time-varying channel environments. The CLEAR system was trained and optimized under specific channel SNR and compression ratio conditions, and its performance was evaluated at multiple SNR levels.

As shown in Fig.~\ref{fig:Compare_Complex}, when the Doppler shift is set to 0.5, the model consistently outperforms cases with other $D_s$ values, regardless of the phase noise. However, as $D_s$ deviates from 0.5 toward 0 or 1, the PSNR improves, leading to better image reconstruction quality. Notably, when both $D_s$ and $P_n$ are minimized, the PSNR exhibits the steepest upward trend, indicating superior reconstruction performance under these conditions.

Furthermore, Fig.~\ref{fig:Compare_Complex} demonstrates that with a fixed $D_s$, smaller $P_n$ values yield higher reconstruction quality as $SNR_{test}$ increases, with the PSNR eventually saturating at high $SNR_{test}$. Interestingly, when $D_s$ is set to $0.2$ or $0.8$, the PSNR trends are nearly identical, consistently surpassing the performance when $D_s$ takes intermediate values (e.g., $0.5$). This phenomenon, which we term the ``moderate degradation effect,'' suggests that the interaction between Doppler shifts and phase noise is not a straightforward linear accumulation. Instead, these interactions result in a nonlinear bimodal relationship due to the complex dynamics of the time-varying channel environment.

These findings highlight the intricate dependencies between Doppler shifts and phase noise on image reconstruction quality. Further exploration of this nonlinear behavior is planned for future work to better understand and mitigate its effects in practical applications.
\begin{figure}[t]
    \centering
     \includegraphics[width=0.4\textwidth]{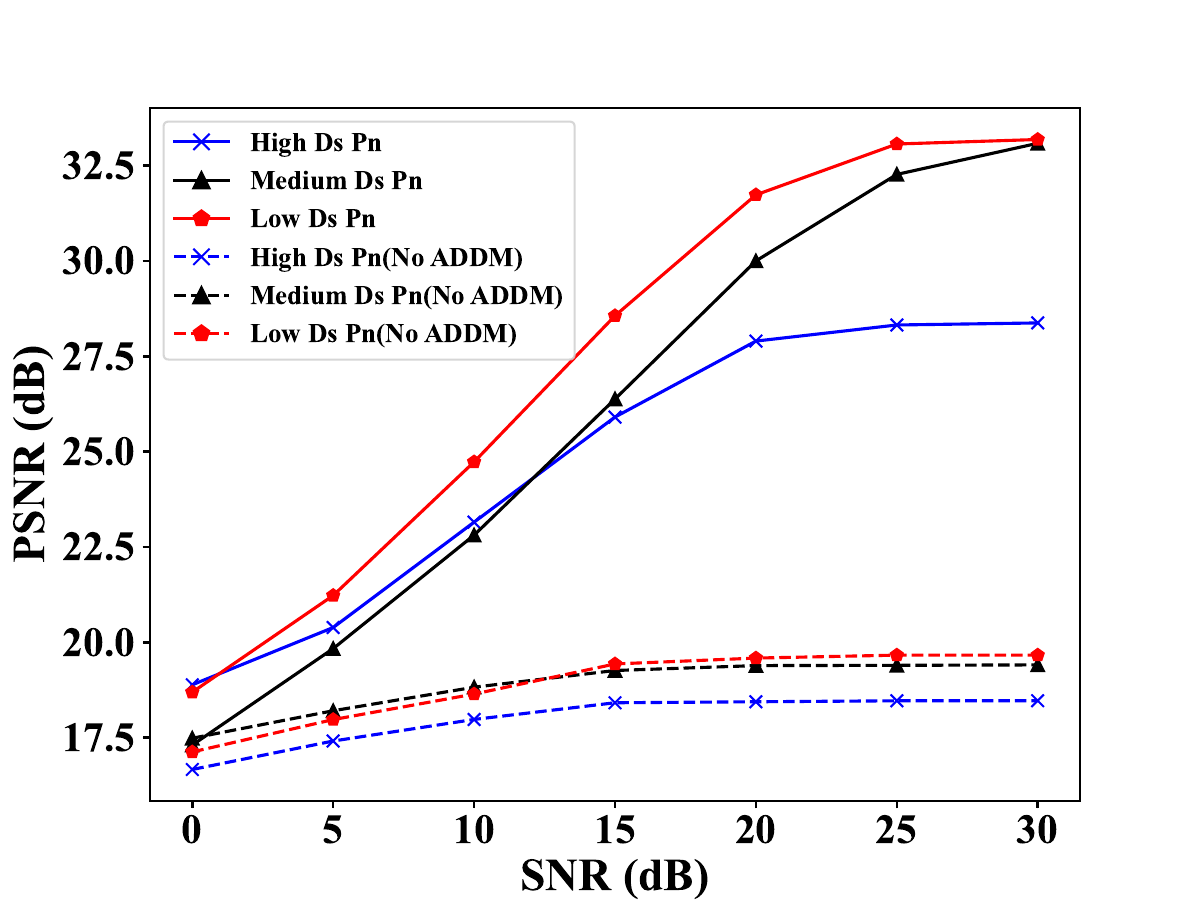}
    \caption{The effect of ADDM denoising on enhancing CLEAR system performance under complex channel conditions.}
    \label{fig:ablation_test}
\end{figure}

\begin{figure*}[t]
    \centering
    \includegraphics[width=5.35in]{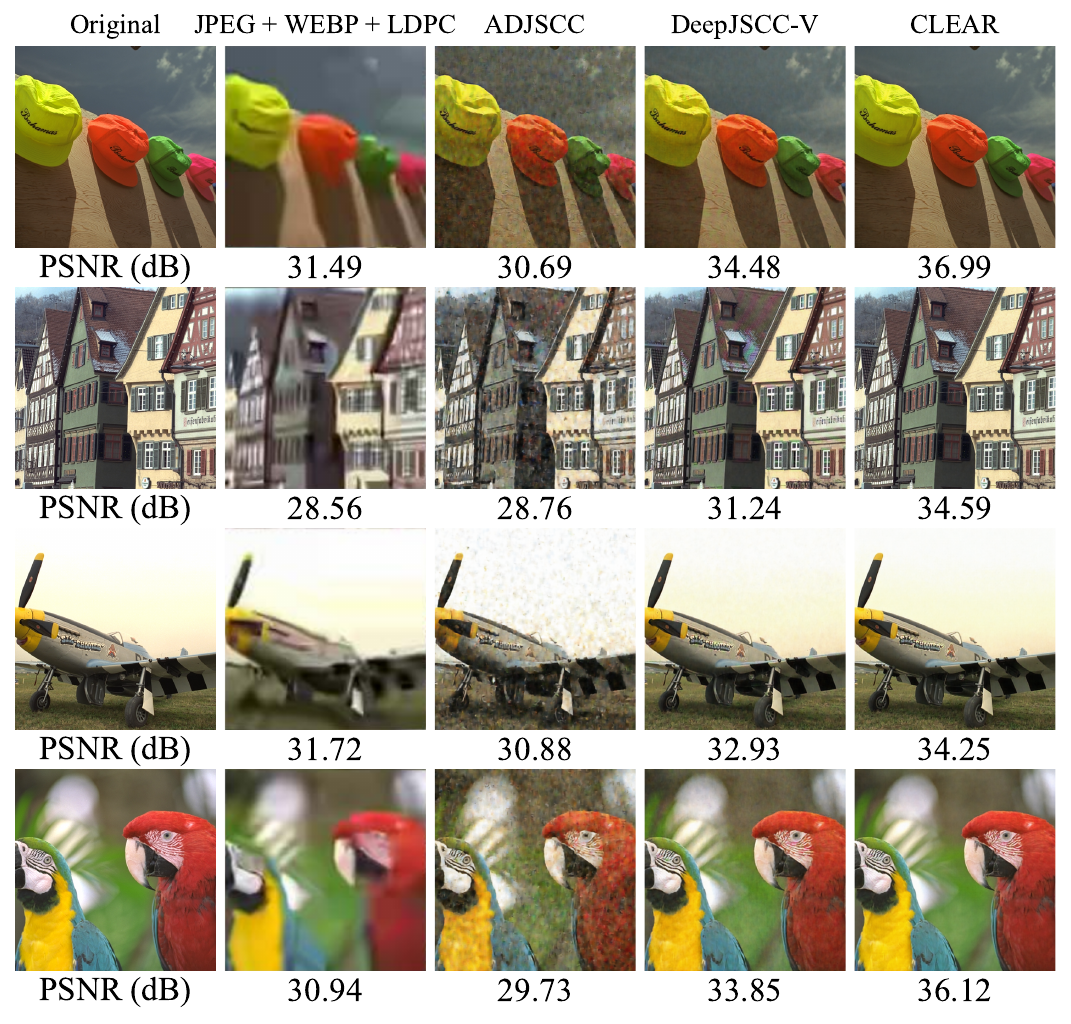}
    \caption{Visual comparison of different transmission schemes in AWGN channel with SNR being set at 15~dB.}
\label{fig:image_test}
\end{figure*}

\vspace{-2pt}
\subsection{Ablation Experiment}
To evaluate the significance of the ADDM denoising module in the CLEAR system, we conducted a comparison of PSNR performance with and without the adaptive diffusion denoising process under complex time-varying channels at varying SNR levels. The results, presented in Fig.~\ref{fig:ablation_test}, highlight the substantial performance improvements achieved with ADDM.

As shown in Fig.~\ref{fig:ablation_test}, the average PSNR of the CLEAR system is consistently higher than that of the model without ADDM. For example, when the SNR is set to $20$~dB, the inclusion of ADDM results in PSNR gains of $9.46$~dB, $10.61$~dB, and $12.15$~dB in high, medium, and low noise channel conditions, respectively. In contrast, without the ADDM denoising process, the system exhibits only gradual PSNR improvements as SNR increases, with the visual quality of reconstructed images remaining suboptimal.

These results demonstrate that the integration of ADDM effectively suppresses noise and preserves image details, significantly enhancing the robustness and overall performance of CLEAR in complex channel environments. By leveraging ADDM, the system achieves superior reconstruction quality, even under challenging conditions characterized by high Doppler shifts and phase noise.

\subsection{Visualization}
To further highlight the effectiveness of the CLEAR system, we present visual results on the Kodak24~\cite{Kodak24} dataset. The comparison was conducted in an AWGN channel with an SNR of 15dB, as shown in Fig.~\ref{fig:image_test}. CLEAR consistently delivers superior visual quality compared to traditional methods such as JPEG+WEBP+LDPC~\cite{Choi2019}, ADJSCC~\cite{Xu2021ADJSCC}, and DeepJSCC-V~\cite{Zhang2023AFB}.

CLEAR achieves an average PSNR gain of approximately $2.3$~dB over DeepJSCC-V. Notably, it effectively suppresses artifacts prevalent in ADJSCC, providing high-fidelity reconstructions with sharper edges and finer details than DeepJSCC-V. The output images of CLEAR are significantly closer to the original, maintaining vivid colors, clearer textures, and a higher level of perceptual realism.

This performance demonstrates the capability of CLEAR to handle complex noise in challenging channel environments while meeting the stringent demands of human visual perception. The results validate the effectiveness of CLEAR as a robust solution for semantic image reconstruction tasks.

\section{Conclusion}
In this paper, we proposed CLEAR, a novel semantic communication system that integrated the strengths of ADDM and DeepJSCC. By dynamically adjusting noise parameters using real-time CSI, the system ensured high-precision transmission while minimizing semantic loss, significantly enhancing robustness and performance in complex, time-varying channel environments. The framework of CLEAR was introduced, including an in-depth analysis of diffusion noise addition and sampling denoising algorithms of ADDM, as well as the training strategies for the entire system. Simulation results demonstrated that CLEAR consistently outperformed traditional methods and existing semantic communication frameworks under identical channel conditions and compression rates, particularly excelling in robustness, noise suppression, and adaptability to complex channel variations. This system showed great potential for application in future wireless communication technologies. Future research directions could include advanced channel modeling, optimization of neural network architectures, and extensions to multimodal communication and secure data transmission, paving the way for innovative advancements in semantic communication systems tailored to dynamic wireless environments.

\appendices
\section{Proof of the Forward Diffusion Process}
As $t$ changes, $x_t$ eventually converges to pure noise. The forward diffusion process 
\begin{equation}
    x_t = \int_{-\infty}^{\infty} h(t, \tau) x_0(t - \tau) d\tau + \sqrt{1 - \alpha_t} \epsilon,
\end{equation}
simplifies to
\begin{equation}
    x_t = \sqrt{\alpha_t} \int_{-\infty}^{\infty} h(t, \tau) x_0(t - \tau) d\tau + \sqrt{1 - \alpha_t} \epsilon_t.
\end{equation}
Assuming that $\alpha_t \to \alpha_T$, it further reduces to 
\begin{equation}
    x_T = \sqrt{\alpha_T}  \int_{-\infty}^{\infty} h(T, \tau) x_0(T - \tau) d\tau + \sqrt{1 - \alpha_T} \epsilon_T.
\end{equation}
When $\alpha_T \to 0$, we obtain
\begin{equation}
    x_T = \sqrt{0} \int_{-\infty}^{\infty} h(T, \tau) x_0(T - \tau) d\tau + \sqrt{1} \epsilon_T = \epsilon_T,
\end{equation}
indicating that $x_T$ is pure noise.

\section{Proof of the Reverse Denoising Process}
The reverse process gradually removes noise and restores the signal. Starting with the forward process, the noisy signal at time step $t$ is represented 
\begin{equation}
    x_t = \sqrt{\alpha_t} \int_{-\infty}^\infty h(t, \tau) x_0(t - \tau) d\tau + \sqrt{1 - \alpha_t} \epsilon_t.
\end{equation}

During the reverse process, the noise is reduced at each step, of which the reverse updating rule is expressed as
\begin{equation}
    x_{t-1} = \frac{1}{\sqrt{\alpha_t}} \left( x_t - \sqrt{1 - \alpha_t} \epsilon_t \right) + \sigma_t \epsilon_t',
\end{equation}
where $\epsilon_t' \sim N(0, I)$ represents the new noise during denoising. To incorporate the effects of the wireless channel, substituting $x_t$ yields
\begin{equation}
    x_{t-1} = \frac{1}{\sqrt{\alpha_t}} 
    \left( \sqrt{\alpha_t} \int_{-\infty}^\infty h(t, \tau) x_0(t - \tau) d\tau \right) + \sigma_t \epsilon_t'
\end{equation}
which simplifies to 
\begin{equation}
    x_{t-1} = \int_{-\infty}^\infty h(t, \tau) x_0(t - \tau) d\tau + \sigma_t \epsilon_t'.
\end{equation}

As the reverse process iterates, the noise $\sigma_t \epsilon_t'$ diminishes progressively due to the decreasing values of $\sigma_t$. By the end of the reverse process, when $t \to 0$, the denoised signal converges to
\begin{equation}
    x_0 = \frac{\lim_{t \to 0} \left( x_t - \sqrt{1 - \alpha_t} \epsilon_t\right)}{\sqrt{\alpha_t}}.
\end{equation}

At this stage, the noise components vanish, and the effective signal components dominate, resulting in the recovery of the original clean signal $x_0$.

\bibliographystyle{IEEEtran}
\bibliography{reference}

\end{document}